\documentclass[12pt]{iopart}
\usepackage{amssymb}	
\usepackage{notes2bib}
\usepackage{threeparttable}
\usepackage{dcolumn}

\newcommand{\ket}[1]{\left| #1 \right>} 
\newcommand{\matrixel}[3]{\big< #1 \big| #2 \big| #3\big>} 
\newcommand{\spinT}[2]{\left| #1 \right> \leftrightarrow \left| #2\right>} 

\newcommand{\wQspin}{\nu_{_Q}}

\newcommand{\HQorb}{\hat{\mathcal{H}}^{\rm{orb}}}

\let\baraccent=\= 
\renewcommand{\=}[1]{\stackrel{#1}{=}} 
\newcommand{\degreeC}{^{\circ}\mathrm{C}} 

\usepackage{graphicx}
\usepackage{dcolumn}
\usepackage{bm}
\usepackage{color}

\newcommand{\onlinecite}[1]{\cite{#1}}

\begin{document}


\title{Quadrupole Shift of Nuclear Magnetic Resonance of Donors in Silicon at Low Magnetic Field}

\author{%
P.~A.~Mortemousque$^{1}$\footnote{Present address: CNRS, Institut NEEL, F-38042 Grenoble, France} \footnote{Corresponding author: pierre-andre.mortemousque@neel.cnrs.fr},
S.~Rosenius$^{1}$,
G.~Pica$^{2}$\footnote{Center for Neuroscience and Cognitive Systems, Italian Institute of Technology, Corso Bettini 31 - 38068 Rovereto, Italy},
D.~P.~Franke$^{3}$,
T.~Sekiguchi$^{1}$\footnote{Present address: Institute for Integrated Cell-Material Sciences (WPI-iCeMS), Kyoto University, Yoshida-Honmachi, Sakyo-ku, Kyoto 606-8501, Japan},
A.~Truong$^{1}$\footnote{Present address: CEA, INAC-SX, F-38000 Grenoble, France},
M.~P.~Vlasenko$^{4}$,
L.~S.~Vlasenko$^{4}$,
M.~S.~Brandt$^{3}$,
R.~G.~Elliman$^{5}$
and
K.~M.~Itoh$^{1}$
}

\address{$^{1}$~School of Fundamental Science and Technology, Keio University, 3-14-1 Hiyoshi, Kohoku-ku, Yokohama 223-8522, Japan}
\address{$^{2}$~SUPA, School of Physics and Astronomy, University of St Andrews, KY16 9SS, United Kingdom}
\address{$^{3}$~Walter Schottky Institut, Technische Universit\"at M\"unchen, Am Coulombwall 4, 85748 Garching, Germany}
\address{$^{4}$~A. F. Ioffe Physico-Technical Institute, Russian Academy of Sciences, 194021, St.~Petersburg, Russia}
\address{$^{5}$~Australian National University, Research School of Physics and Engineering, Canberra,  ACT 0200, Australia}

\begin{abstract}
Shifts from the expected nuclear magnetic resonance frequencies of antimony and bismuth donors in silicon of greater than a megahertz are observed in electrically detected magnetic resonance spectra. 
Defects created by ion implantation of the donors are discussed as the source of effective electric field gradients generating these shifts via quadrupole interaction with the nuclear spins. 
The experimental results are modeled quantitatively by molecular orbital theory for a coupled pair consisting of a donor and a spin-dependent recombination readout center.
\end{abstract}

\maketitle


Group-V donors are the most studied impurities in silicon (Si). 
For decades, the shallow donors have been employed for the fabrication of transistors in microelectronics. 
More recently, they have become very promising candidates as quantum bits \cite{Kane1998}, since they can be considered as  hydrogen-like atoms embedded in a solid-state matrix. 
The weak spin-orbit coupling present in the silicon host makes the coherence time of the donor electron and nuclear spins exceptionally long \cite{Ladd2005, Tyryshkin2006}.   
Moreover, the existence of spin-free $^{28}$Si allows for the isotopic purification of Si \cite{Takyu1999, Andreas2011, Itoh2014},  further improving these coherence times by removing most of the $^{29}$Si (nuclear spin $I=1/2$) \cite{Abe2010, Tyryshkin2012, Steger2012, Wolfowicz2013, Saeedi2013}. 
This has motivated researchers to investigate techniques to incorporate donors in silicon to produce single donor devices \cite{Morello2010, Pla2013, Muhonen2014}.  
The electron and nuclear spin $g$-factors of shallow donors are already well-known \cite{Feher1959}.  
Similarly, the hyperfine interaction between the donor electron with donor nuclear spin \cite{Feher1959, Wilson1961} and with $^{29}$Si nuclear spins \cite{Hale1969} have been measured and are used to understand the donor electron wavefunction \cite{Faulkner1969, Ivey1975}.  More recently, the Stark shift of the donor states has been observed \cite{Bradbury2006, Lo2014, Wolfowicz2014} and modeled \cite{Smit2004, Friesen2005, Rahman2007, Rahman2009, Pica2014}. 

Beyond their coherence properties, group-V donors in silicon with large nuclear spins ($> 1/2$)  such as $^{75}$As, $^{121}$Sb, $^{123}$Sb, and $^{209}$Bi (see Table~\ref{Table:SpinProperties}) provide additional accessible states which are very attractive resources for performing quantum computations. 
Muthukrishnan and Stroud \cite{Muthukrishnan2000} have shown that the use of qudits \cite{Neeley2009}, $d$-level systems, allows to decrease the number of necessary elementary qubits for a computation by a factor of $\log_2{d}$. 
Moreover, Lanyon \textit{et al.} \cite{Lanyon2009} have demonstrated that the five two-qubit gates required to realize a Toffoli gate (a three-qubit entangling gate) can be reduced to only three gates if at least one of the three qubits has an additional available level. 
Also, an implementation of the Grover algorithm directly in large nuclear spins in semiconductors has been proposed by Leuenberger and Loss \cite{Leuenberger2003-2}.

This paper reports the experimental observation of line shifts of the nuclear spin resonance of $^{121}$Sb, $^{123}$Sb and $^{209}$Bi donors in Si at low magnetic fields observed via the electrical detection of mixed spin states. 
These shifts are interpreted as resulting from the nuclear quadrupole interaction (abbreviated NQI) of the nuclear electric quadrupole moment with the donor electron wavefunction, which is modified by the presence of implantation defects.
A shift of the nuclear magnetic resonance (NMR) due to NQI arises from the interaction of a nuclear spin $I\geq1$ with an electric field gradient (EFG) and is well-studied in the atomic physics and nuclear magnetic resonance communities \cite{Kaufmann1979, Jeon1989, Kotegawa2015}. 
However, the observation of a quadrupole-induced magnetic resonance shift of the group-V donors in Si, where the NQI with crystal field gradients vanishes due to the cubic symmetry, has only recently been realized for ionized \cite{Franke2015} and neutral \cite{Franke2016} donors in strained Si, and for $^{209}$Bi implanted silicon devices \cite{Pla2016}. 
While an external application of EFG large enough to result in an observable shift ($\gtrsim10^{18}$~V/m$^2$) is challenging, gradients of such a magnitude can be generated by donor electrons whose spatial symmetry is broken by the presence of strain \cite{Ishihara2014}, close acceptors \cite{Burchard1999} or ion implantation defects \cite{Elkin1968, Watkins1997, Larsen2006}. 
As our detection method is based on the spin-dependent recombination (SDR) \cite{Lepine1972, Mortemousque2012, Mortemousque2014} of close pairs ($1\sim10$ nm) formed by the donor electron and such defects, only donors that are exposed to these EFG are probed. By working at low magnetic fields, where the donor electron and nuclear spin wavefunctions mix strongly due to the hyperfine interaction, we are able to use such SDR signal to investigate possible quadrupole-induced NMR shifts \cite{Mortemousque2012}.

Three samples, Si:$^{121}$Sb, Si:$^{123}$Sb, and Si:$^{209}$Bi, were employed. 
The substrate used for fabrication was from a highly resistive ($>3$ k$\Omega \cdot$cm) float-zone Si wafer. 
The samples were ion-implanted with $^{121}$Sb, $^{123}$Sb or $^{209}$Bi at room temperature. 
Bi was implanted at 300 and 550 keV with doses of $0.7\times10^{13}$ and $1.3\times10^{13}$ cm$^{-2}$, respectively. 
The isotope-selective Sb implantations were performed with an energy of 30~keV and a fluence of $1.7\times10^{11}$ cm$^{-2}$.
These conditions yielded a maximum Bi plateau-like concentration of $1.8\times10^{18}$ cm$^{-3}$ [above the solubility limit \cite{Trumbore1959}] in a depth of 90 to 150 nm from the surface, and a $^{121,\,123}$Sb peak concentration of $1.1\times10^{17}$ cm$^{-3}$ 24 nm from the surface. 
The three samples were individually annealed after the ion implantation, at 650 $\degreeC$ for 30 min in an evacuated quartz tube. 
This process, designed to maximize the number of donor-readout center (D-R) pairs \cite{Studer2013}, led to an activation efficiency \cite{Marsh1968,Baron1969,deSouza1993,Weis2012}  $\lesssim60\%$, resulting in a Bi donor concentration less than $1.1\times10^{18}$ cm$^{-3}$, and a $^{121}$Sb and a Si:$^{123}$Sb concentration less than $4\times10^{16}$ cm$^{-3}$ (in all cases below the metal-insulator transitions \cite{Ochiai1975, Abramof1997}). 

%
%
\begin{figure}[b]
\centering
\includegraphics[width=100mm]{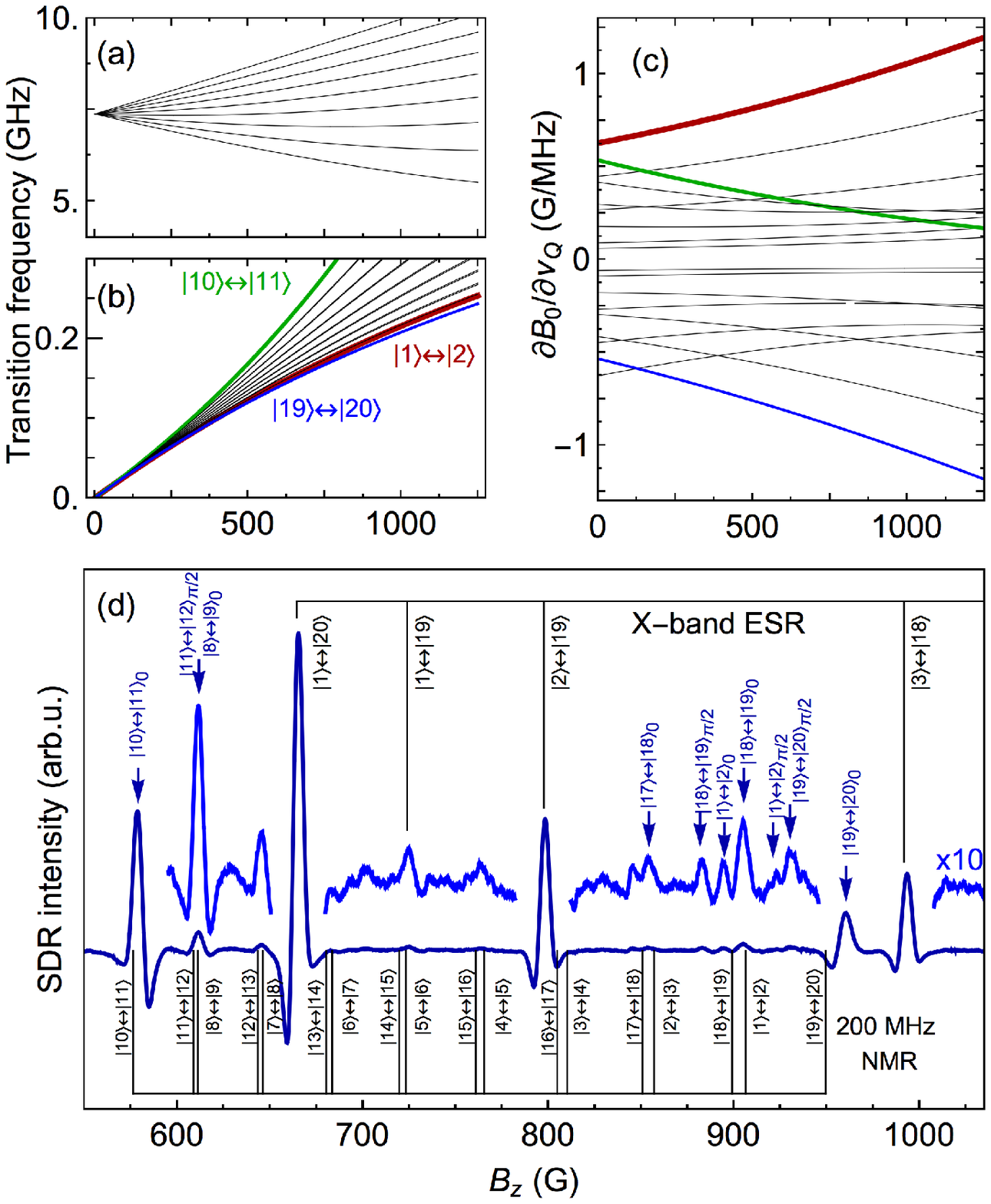}
\caption{\label{Fig_NMRBi}
(Color online.) 
(a) ESR and (b) NMR transition frequencies and (c) NMR resonant magnetic field sensitivity to the NQI of Bi donors as a function of magnetic field. 
The colored lines ($\spinT{19}{20}$ in blue, $\spinT{10}{11}$ in green and $\spinT{1}{2}$ in red) have high SDR intensities and were analysed in this study. 
The same color code is used in Fig.~\ref{Fig:shiftVSsensitivityBi}(a, b and d). 
(d) Experimentally observed SDR spectrum of the Bi donor spin transitions under 200-MHz excitation. 
The black lines represent the expected resonance position of the electron and nuclear spin transitions without exchange interaction or NQI. 
The arrows indicate the assignment of the spin transitions between the states $\ket{i}$ and $\ket{j}$, for an angle $\theta = 0$ or $\theta = \pi/2$ (see text). 
}
\end{figure}

In the SDR detection technique, the photocarriers generated by a 100-W halogen lamp are captured by the ionized donors. 
The recombination time of the donor electrons depends on the relative electron spins of the donor and the paramagnetic readout center: in this D-R pair, the antiparallel electron spin configuration leads to a fast recombination, whereas for the parallel spins, the expected recombination time is much longer (spin blockade) \cite{Hoehne2013-2}. 
In a steady state and for a continuous illumination, mostly the parallel electron spin pairs remain without external induction of magnetic resonance. 
Therefore, flipping the donor electron spins by magnetic resonance enhances the electron recombination, leading to a decrease in the sample photoconductivity from the steady state. 
This change in photoconductivity was measured by probing the change in the microwave reflectivity of the sample in a JEOL JES-RE3X X-band cw-electron spin resonance (ESR) spectrometer tuned at 9.086~GHz and 100~mW,  with an Oxford $^4$He-cryostat and a lock-in amplifier using a modulation amplitude of 7~G at 100~kHz. 
For the low frequency excitations, a dedicated coil was used. 
A static magnetic field is applied in the crystallographic [001] direction.

\Table{\label{Table:SpinProperties} 
Magnetic resonance parameters of group-V $^{75}$As, $^{121}$Sb, $^{123}$Sb and $^{209}$Bi 
donors in Si. 
The quadrupole moment of the donor nuclei $\mathcal{Q}$ is given in $10^{28}$~m$^{-2}$. 
$\bar{a}_l$ is the Bohr radius as defined in \cite{Pica2014}.}
\br
& Si:$^{75}$As & Si:$^{121}$Sb & Si:$^{123}$Sb & Si:$^{209}$Bi \\
\mr
$I$ & 3/2 & 5/2 & 7/2 & 9/2 \\
$g_e$ & 1.99837$^{\rm a}$& \multicolumn{2}{c}{1.9986$^{\rm a}$} & 2.00049(5)$^{\rm b}$ \\
$g_n$ & 0.9599$^{\rm e}$ & 1.3454$^{\rm e}$ & 0.7285$^{\rm e}$ & 0.9135$^{\rm e}$ \\
$A$ (MHz)& 198.35$^{\rm a}$ & 186.80$^{\rm a}$ & 101.52$^{\rm a}$ & 1475.05(17)$^{\rm b}$ \\
$\bar{a}_l$ (nm) & 1.45$^{\rm d}$ & \multicolumn{2}{c}{1.67$^{\rm d}$} & 0.967$^{\rm d}$ \\
$\mathcal{Q}$ (b) & $+0.307(4)^{\rm e}$  & $-0.41(4)^{\rm e}$ & $-0.49(5)^{\rm e}$ & $-0.58(9)^{\rm e}$\\
\br
\end{tabular}
\item[] $^{\rm a}$ Feher (Ref. \onlinecite{Feher1959})
\item[] $^{\rm b}$ Mortemousque \textit{et al.} (Ref. \onlinecite{Mortemousque2014})
\item[] $^{\rm c}$ Wolfowicz \textit{et al.} (Ref. \onlinecite{Wolfowicz2013})
\item[] $^{\rm d}$ Pica \textit{et al.} (Ref. \onlinecite{Pica2014})
\item[] $^{\rm e}$ Stone (Ref. \onlinecite{Stone2005})
\end{indented}
\end{table}
%
%

The donor spin Hamiltonian considered is 
\begin{equation}
\label{Eq:SpinHamiltonian}
\hat{\mathcal{H}}=g_e \mu_B B_z \hat{S_z} - g_n \mu_N B_z \hat{I_z} + h A \,\vec{\hat{S}}\cdot\vec{\hat{I}} + \vec{\hat{I}}\cdot Q \cdot\vec{\hat{I}},
\end{equation}
where $g_e$ and $g_n$ are the donor electron and nuclear $g$-factors, respectively, $A$ is the strength of the isotropic hyperfine interaction (Table \ref{Table:SpinProperties}), and $B_z$ is the magnetic field in the $z$ direction. 
The eigenstates $\ket{i}$ of Eq.~(\ref{Eq:SpinHamiltonian}) are $\cos \phi_i \ket{m_S = 1/2, m_I} + \sin \phi_i \ket{m_S = 1/2, m_I+1}$, where $m_I$ and $m_S$ are the eigenvalues of $\hat{I_z}$ and $\hat{S_z}$, and $\phi_i$ is the spin mixing angle, explicitly described in Ref.~\cite{Mohammady2010}. 
They are labeled with integers $1\leq i \leq (2S+1)(2I+1)$ in order of increasing energy. 
The term  $\vec{\hat{I}}\cdot Q\cdot\vec{\hat{I}}$ represents the quadrupole interaction of the donor nuclear spin. 
For an electrostatic field $E_z$ experienced by the nucleus in presence of the electronic charge configuration of its atom, 
the effective EFG in $z$ direction is $\partial E_z/\partial z = V_{zz}^{\rm{eff}}$, and the NQI is written as \cite{Kaufmann1979} 
\begin{equation}
\label{Eq:QuadrupoleHamiltonian}
\vec{\hat{I}}\cdot Q\cdot\vec{\hat{I}} = h\wQspin \left(3\hat{I_z}^2-I(I+1)\right)/(4I(2I-1)),
\end{equation}
where the coefficient $h\wQspin=e\mathcal{Q}V_{\rm{zz}}^{\rm{eff}}$, $e$ is the elementary electrical charge and $\mathcal{Q}$ is the quadrupole moment of the nucleus.

At high magnetic fields, the eigenstates of Eq.~(\ref{Eq:SpinHamiltonian}) are given by the products of the electron and nuclear spin states. 
In this case, their magnetic resonances are addressed separately, either in ESR ($\Delta m_I=0$, $\Delta m_S=\pm 1$) or NMR ($\Delta m_S=0$, $\Delta m_I=\pm 1$). 
In SDR spectroscopy, which is sensitive to the electron spin configuration, only terms involving the electron spin operator $\vec{\hat{S}}$ can be detected, namely the electron Zeeman interaction and the hyperfine interaction.
For $hA\gtrsim g_e\mu_B B_z$, however, the eigenstates are linear combinations of several nuclear and electron states, so-called mixed states \cite{Morishita2009, Dreher2015}. 
Resonant transitions which change both spin states become possible, allowing for transitions which are sensitive to the nuclear spin properties but also change the electron spin state to become detectable by SDR \cite{Mortemousque2012, Franke2014}. 
The sensitivity of the low-field line positions (resonant magnetic fields) to the NQI, noted $\partial B_0/\partial\nu_Q$, were calculated numerically for $^{209}$Bi [Fig.~\ref{Fig_NMRBi}(c)].
The spin-dependent conductivity signal is based on coupled spin pairs, which are most efficiently formed for `pure' electron spin states \cite{Morishita2009, Mortemousque2012}. 
The transitions most easily detected [Fig.~\ref{Fig_NMRBi}(b, c)] are therefore those involving the two product states: $\ket{m_S = + S}\ket{m_I = + I}=\ket{20}$ ($\spinT{19}{20}$, in blue) and $\ket{m_S = - S}\ket{m_I = - I}=\ket{10}$ ($\spinT{10}{11}$, in green). 
Other NMR transitions have a weaker SDR intensity [Fig.~\ref{Fig_NMRBi}(d)]. 

The ESR and NMR transition frequencies of Bi donors are plotted in Fig.~\ref{Fig_NMRBi}(a, b). 
Figure~\ref{Fig_NMRBi}(d) shows an SDR spectrum of $^{209}$Bi donors in Si recorded under a 200-MHz excitation. 
Four of the observed peaks are identified as the resonance transitions induced by the X-band microwave employed for the SDR detection of the ac-conductivity of the sample. 
They can be used to extract the electron $g$-factor and the hyperfine interaction $A$ since they are, in good approximation, not influenced by the nuclear Zeeman interaction or the NQI. 
As discussed in Ref.~\onlinecite{Mortemousque2014}, the observed values are slightly different from those observed in conventional ESR due to the influence of the recombination partner. 
In addition, several peaks are observed due to the rf excitation, in particular two prominent lines at $\approx 575$~G and $\approx 960$~G attributed to the NMR transitions $\spinT{10}{11}$ and $\spinT{19}{20}$, respectively. 
However, when comparing these to the line positions as calculated for Eq.~(\ref{Eq:SpinHamiltonian}), using the experimental values of $A$ and $g_e$ extracted from the X-band line positions and assuming zero NQI [Fig.~\ref{Fig_NMRBi}(d)], the NMR lines appear to be shifted significantly. 
These line shifts could not be observed in our previous report \cite{Mortemousque2012} due to the lower concentration in Bi donors. 
We could not observe forbidden electron spin transitions (except $\spinT{1}{19}$) or spin transitions arising from mixed mw-RF excitation frequencies.
Octupole interaction was not observed and higher order multipole interactions are assumed to be negligible.

%
%
\begin{figure}[!t]
\centering
\includegraphics[width=100mm]{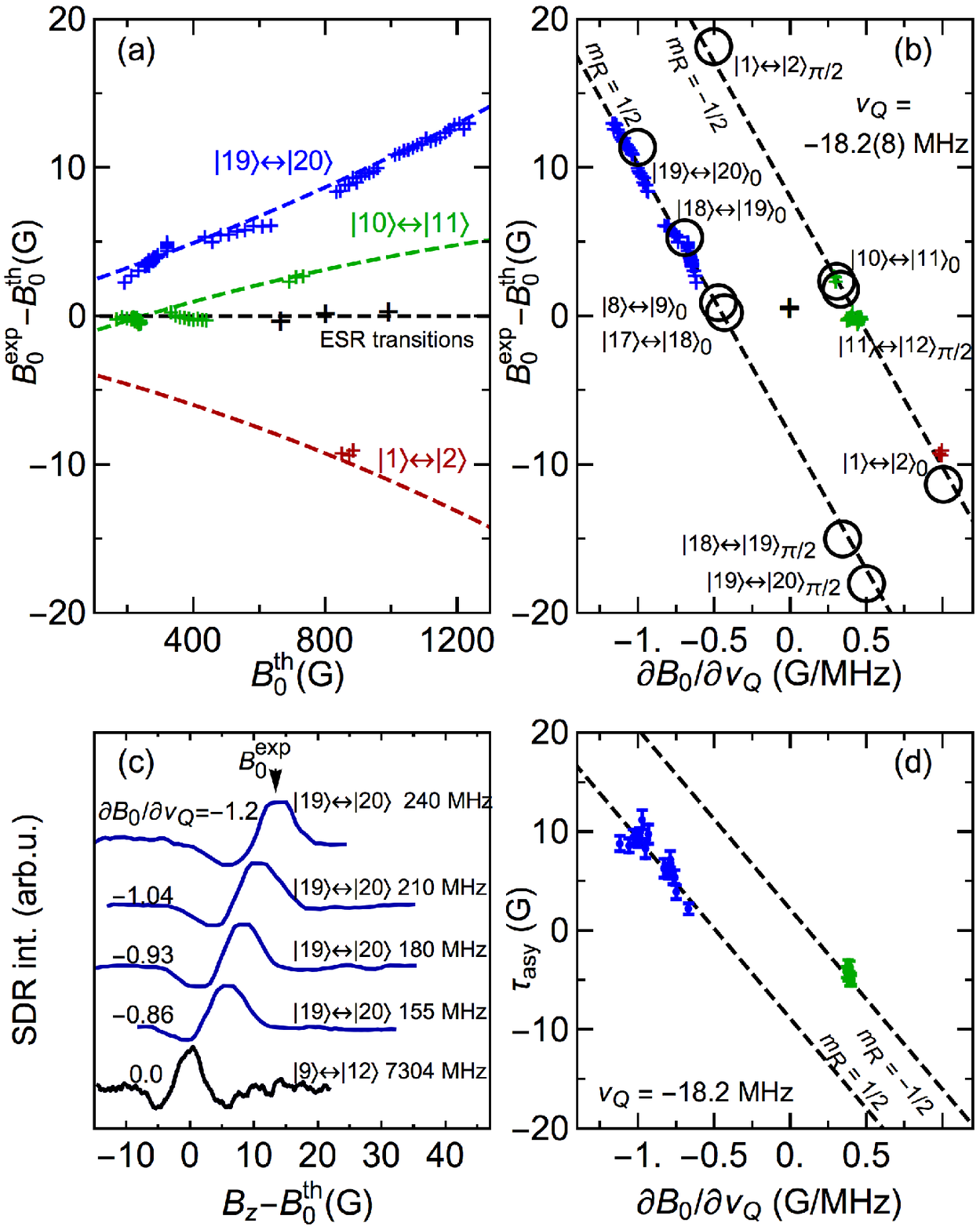}
\caption{\label{Fig:shiftVSsensitivityBi}
(Color online.)
Resonant magnetic field shift $B_0^{\rm{exp}}-B_0^{\rm{th}}$ of $\spinT{19}{20}$ (blue crosses),  $\spinT{10}{11}$ (green crosses), and $\spinT{1}{2}$ (red crosses) of Si:$^{209}$Bi plotted as a function of $B_z$ (a) and the sensitivity $\partial B_0/\partial \wQspin$ (b). 
The circles represent the resonant magnetic field shifts estimated from Fig.~\ref{Fig_NMRBi}(d), but not taken into account for the fit of the NQI and exchange interaction parameters. 
The black crosses correspond to the $\spinT{1}{20}$, $\spinT{2}{19}$ and $\spinT{3}{18}$  X-band ESR transitions.  
$B_0^{\rm{exp.}}$ and $B_0^{\rm{th}}$ are the measured and calculated resonant fields (assuming that $\wQspin = 0 $).
The dashed lines are calculated from a single linear fit (see text), resulting in $\wQspin=-18.2(2)$~MHz and an exchange interaction $J=45(2)$~MHz. 
(c) Recorded SDR spectra of the Bi spin transitions $\spinT{19}{20}$ (blue) measured at various frequencies. 
The spectrum (black) of the $\spinT{9}{12}$ transition at the hyperfine clock transition \cite{Mortemousque2014} is plotted for comparison (insensitive to distribution in hyperfine interaction, and very weakly sensitive to NQI).  
(d) Asymmetry parameter $\tau_{\rm{asy}}(\partial B_0/\partial \wQspin)$ (see text) plotted for Si:Bi spectra (with high enough signal to noise ratio and non overlapping transition lines).  
The dashed lines (a and d) are guides for this eye and were calculated assuming $\wQspin=-18.2$ MHz and $J=45$~MHz. }
\end{figure}

To gain further insight into the origin of the observed shift, 
the resonance lines are recorded at different excitation frequencies as shown in Fig.~\ref{Fig:shiftVSsensitivityBi}(c), where the horizontal axes are aligned to the expected resonance field $B_0^{\rm{th}}$ for the spin system without NQI. 
For all the measured NMR spectra of Si:$^{209}$Bi, the shift of the resonant magnetic field for three transitions is shown in Fig.~\ref{Fig:shiftVSsensitivityBi}(a) as a function of the magnetic field. 
Since the observed shifts are orders of magnitude larger than could be expected for a change in nuclear or electron $g$-factor, or in hyperfine interaction, we will now discuss the NQI as a possible explanation.
To separate the effect of the NQI from other possible influences, the data are plotted against 
$\partial B_0/\partial\nu_Q$ of the corresponding transitions [Fig.~\ref{Fig:shiftVSsensitivityBi}(b)]. 
The line positions extracted from the $\spinT{19}{20}$, $\spinT{10}{11}$ and $\spinT{1}{2}$ transitions are well described by a linear dependence, which confirms the NQI of the nucleus as origin of the observed shift. 
The linear fit slopes of Fig.~\ref{Fig:shiftVSsensitivityBi}(b) indicates a NQI $\wQspin$$^{\rm{exp}}_{\rm{Bi}} = -18.2(2)$ MHz in Si:$^{209}$Bi. 
The corresponding effective EFG is $V_{zz}^{\rm{eff,Bi}} = 1.30\times 10^{21}$~V/m$^{2}$. 

However, an additional offset is necessary to reproduce the observed line shifts. 
A possible explanation is the exchange interaction ($hJ\,\vec{\hat{S_{_R}}}\cdot\vec{\hat{S}}$) between the readout partner (electron spin state $m_{_R}$) and the donor electron spins which was not considered in Eq.~(\ref{Eq:SpinHamiltonian}), and leads to a constant shift of the resonance for the transitions and frequency region considered here. 
This is supported by the different offset signs for the three transitions, in agreement with the steady state polarization of the spin pairs due to the spin-dependent recombination process. 
With the expected sensitivity $\partial B_0 / \partial J = -m_{_R}\frac{h}{g_e\,\mu_{_B}} \approx -0.18$~G/MHz for $\spinT{19}{20}$ ($+0.18$ for $\spinT{10}{11}$ and $\spinT{1}{2}$), the observed $\mp8.0(4)$ G offsets in Fig.~\ref{Fig:shiftVSsensitivityBi}(b) can be explained by a coupling constant $J(\rm{Bi}) = 45(2)$~MHz, significantly stronger than usually observed in SDR experiments \cite{Suckert2013}. 
Another effect contributing to these offsets is the asymmetric and broad distribution of the NQI (assumed exponential).
Such a distribution makes the resonance line asymmetric (see Fig.~\ref{Fig:shiftVSsensitivityBi}(c)). 
The line shape, symmetric for $\partial B_0 / \partial \wQspin \approx 0$,  becomes very asymmetric at more negative $\partial B_0 / \partial \wQspin$, revealing the NQI distribution. 
More quantitatively, the parameter $\tau_{\rm{asy}}$ of the fitting function $\rmd^2\left\{\exp(-B/\tau_{\rm{asy}})\otimes\exp(-(B-B^{\rm{exp}})^2/2\sigma^2)\right\}/\rmd B^2$ is plotted in Fig.~\ref{Fig:shiftVSsensitivityBi}(d).
%
%
The line offsets in the latter indicate 
that the asymmetry of the resonance peaks is not only caused by NQI, but by exchange interaction as well.

%
%
\begin{figure}[!b]
\centering
\includegraphics[width=100mm]{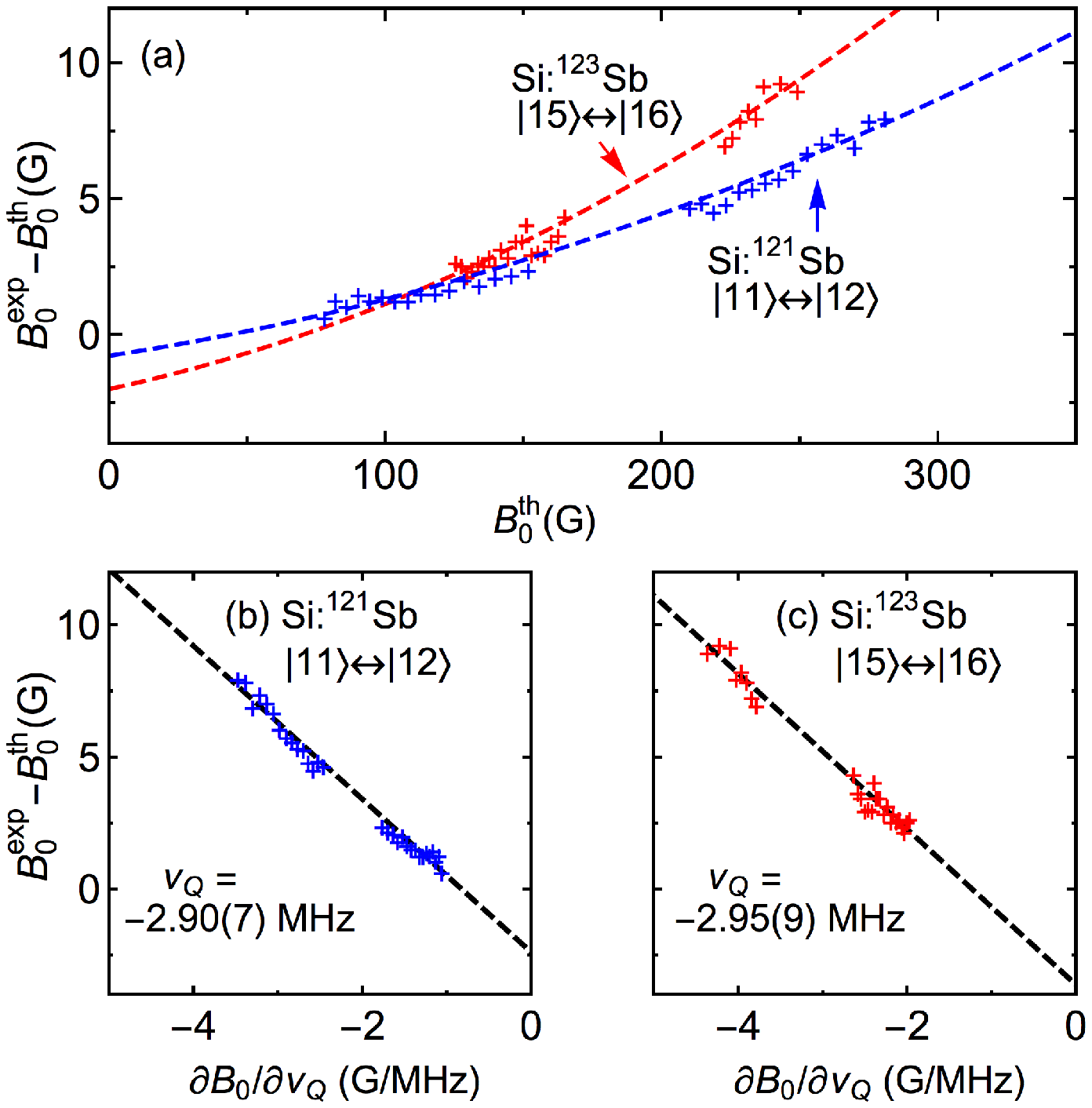}
\caption{\label{Fig:shiftVSsensitivitySb}
(Color online.)
(a) Detuning $B_0^{\rm{exp}}-B_0^{\rm{th}}$ of the transitions $\spinT{11}{12}$ in Si:$^{121}$Sb (blue) and  $\spinT{15}{16}$ in Si:$^{123}$Sb (red) plotted against $B_0^{\rm{th}}$. The corresponding lines are guides for the eye calculated using the fitting parameters. 
(b, c) Shift, as defined in Fig.~\ref{Fig:shiftVSsensitivityBi}, observed for various transitions of Si:$^{121}$Sb (b) and  Si:$^{123}$Sb (c) plotted as a function of $\partial B_0/\partial \wQspin$. 
The linear fits indicate $\wQspin(^{121}\rm{Sb}) = -2.90(7)$~MHz and $\wQspin(^{123}\rm{Sb}) = -2.95(9)$~MHz. 

}
\end{figure}

To investigate further the origin of the EFG responsible for the NQI, we reproduced the same measurement protocol with the two isotopes of the Sb donors in Si, which have an equal Bohr radius, different from Bi donors, allowing us to characterize the role of the donor electron distribution in the generation of the EFG. 
The shift of the resonant magnetic field for the transitions of the two samples are shown in Fig.~\ref{Fig:shiftVSsensitivitySb}(a) as a function of the magnetic field.
Figure~\ref{Fig:shiftVSsensitivitySb}(b) [respectively (c)] shows the measured shift of NMR transitions of $^{121}$Sb [$^{123}$Sb] donors against $\partial B_0 / \partial \wQspin$. 
A linear fit of the points indicates a NQI $\wQspin^{\rm{exp}}(^{121}\rm{Sb})=-2.90(7)$~MHz and $\wQspin^{\rm{exp}}(^{123}\rm{Sb})=-2.95(9)$~MHz, and an exchange interaction of $J(^{121}\rm{Sb}) = 13(2)$~MHz and $J(^{123}\rm{Sb}) = 20(3)$~MHz. 

The quadrupole moment ratio $\mathcal{Q}$$_{^{121}\rm{Sb}}/\mathcal{Q}$$_{^{123}\rm{Sb}}$ is 0.85(17). 
An equivalent ratio is expected for $\wQspin^{\rm{exp}}(^{121}\rm{Sb}) / \wQspin^{\rm{exp}}(^{123}\rm{Sb})$. In this study, the experimental value of this ratio is $0.98(5)$, which is in good agreement with the theoretical value. 
Though further studies could help to quantitatively explain the measured line positions, e.g. with electron nuclear double resonance experiments in the high-field limit, our experiments indicate NQI in the order of several MHz, corresponding to effective EFG on the order of $10^{20}$~V/m$^2$.

The NQI with an uniaxial EFG shows a dependence on the angle $\theta$ between gradient axis and the quantization axis \cite{LevittMAN}, which is in first order $(3\cos^2\theta-1)/2$. 
The tetrahedral symmetry of the donor is broken by the presence of a defect center and for a random distribution of defect sites, one expects a continuous distribution of $e\mathcal{Q} V_{zz}/h$ from $-\nu^{max}_Q/2$ to $\nu_Q^{max}$ for a quantization axis independent from the NQI. 
The lock-in detection of the change in the sample conductivity used in this study allows us to detect only the edge of the broadened resonance peaks. 
Therefore a continuous distribution in $V_{zz}$ leads to an effective peak splitting [blue labels in Fig.~\ref{Fig_NMRBi}(d)]. 
The signal-to-noise ratio achieved experimentally did not allow the observation of all resonance peak splittings and the spectral density of the resonances limited the peak assignment. 

No control of the electric field or of the EFG was performed during the experiments and the EFG induced by local strain \cite{Franke2015} is  negligible as no evidence of strain could be found in Si:Bi  \cite{Mortemousque2014} nor in Si:Sb at X-band.
Therefore, we propose that the EFG at the nucleus is induced by a redistribution of the donor electron density. 
To confirm this hypothesis, we proceeded in two steps. 
First, the orbital of the donor-readout center electron pair can be modeled to evaluate the mixing of the different valleys of the donor 1s ground state. 
This was performed using a variational calculation using a wavefunction basis that includes the six donor valley configurations of the 1s state plus a localized readout center \cite{Mortemousque2014}. 
The energy levels of the donor ground states are known from optical studies \cite{Ramdas1981}, whereas the localised readout center energy level is assumed to be $E_{_R} = -0.25$ eV \cite{Poindexter1984}. 
The envelop functions were calculated using a corrected effective mass theory donor wavefunction. 
The parameter $\eta = 159.4$ described in Ref.~\onlinecite{Assali2011} was used to adjust the envelop functions close to the donor nucleus \cite{Kohn1955}. 
Due to the presence of the readout center, the different valley populations of the donor ground state adjust to  become $\psi = \alpha_{\rm{A1}} \phi_{\rm{A1}} + \sum \alpha_i \phi_i$, where $\phi_i$ are the twofold E and threefold T$_2$ valley configurations. 
The coefficients $\alpha_i$ depend strongly on the relative positions of the two recombination partners with respect to the crystal axes. 


The second step was the computation of the EFG for each donor valley configuration. 
Then, the EFG $V_{x_ix_i}$ generated by an orbital $\psi$  was computed for each donor valley configuration 
as \cite{Kaufmann1979} 
\begin{equation}
\label{Eq:QuadrupoleIntegral}
\matrixel{\psi }{\HQorb_{x_ix_i}}{\psi} = \frac{e}{4\pi\epsilon_0\epsilon_{\rm{Si}}}\int\psi^\ast \frac{3x_i^2-r^2}{r^5}\psi \,\,\rmd\tau,
\end{equation}
where $x_i=x,y,z$. 
According to the group theory, the first order term of the EFG is of the form $\alpha_i\matrixel{\rm{A}_1}{\HQorb}{\phi_i}$, and is significant only for the E-doublet valley configuration (the second and higher order correction terms are negligible, see Table \ref{Table:EFGsummary}).

%
%
\begin{table}
\caption{\label{Table:EFGsummary} 
Simulated values, using Eq.~(\ref{Eq:QuadrupoleIntegral}), of the EFG for different valley configurations for $I\geq1$ group-V donors. Only values of EFG larger than $10^{15}$~V/m$^2$ are listed. 
The EFG numerical values are given in units of $10^{19}$~V/m$^2$. We use the valley notation A$_1$: $(X+\bar{X}+Y+\bar{Y}+Z+\bar{Z})/\sqrt{6}$, E$_{\rm{x,y}}$: $(X+\bar{X}-Y-\bar{Y})/2$, and E$_{\rm{2z}}$: $(2Z+2\bar{Z}-X-\bar{X}-Y-\bar{Y})/\sqrt{12}$. 
}
\begin{indented}
\item[]\begin{tabular}{cD{.}{.}{1}D{.}{.}{1}D{.}{.}{1}}
\br
  & $Si:As$ & $Si:Sb$ & $Si:Bi$ \\
\mr
$\matrixel{\rm{A}_1}{\HQorb_{xx,}}{\rm{E}_{\rm{x,y}}}$  & 3.0 & 2.7 & 11.6 \\
$\matrixel{\rm{A}_1}{\HQorb_{yy,}}{\rm{E}_{\rm{x,y}}}$  & -3.0 & -2.7 & -11.6 \\
$\matrixel{\rm{A}_1}{\HQorb_{zz}}{\rm{E}_{\rm{x,y}}}$  & 0.0 & 0.0 & 0.0 \\
$\matrixel{\rm{A}_1}{\HQorb_{xx}}{\rm{E}_{\rm{2z}}}$  & -1.7 & -1.6 & -6.7 \\
$\matrixel{\rm{A}_1}{\HQorb_{yy}}{\rm{E}_{\rm{2z}}}$  & -1.7 & -1.6 & -6.7 \\
$\matrixel{\rm{A}_1}{\HQorb_{zz}}{\rm{E}_{\rm{2z}}}$  & 3.4 & 3.2 & 13.4\\
\br
\end{tabular}
\end{indented}
\end{table}
%
%

For the SiBi sample, the average D-R separation has been calculated to be 2.3 $\bar{a}_l$ \cite{Mortemousque2014}. The estimated value of $\wQspin^{\rm{th}}(\rm{Bi})$ for such D-R separation is 0.11 MHz, causing an EFG underestimated by a factor $\wQspin^{\rm{exp}}(\rm{Bi})/\wQspin^{\rm{th}}(\rm{Bi})=160$. 
This discrepancy might be explained by the antishielding phenomenon. 
Upon application of an EFG, the energy levels of all shells of an atom are modified, giving rise to an EFG induced by the subshell electrons. 
Feiock and Johnson \cite{Feiock1969} have shown that a first order correction to the subshell energy levels gives the total EFG at the donor nucleus $V_{zz}^{\rm{tot}} = (1-\gamma) V_{zz}^{\rm{eff}}$, where $\gamma$ is the antishielding or Sternheimer coefficient. 
The antishielding parameters of donors in Si have been measured only for limited species and ionization numbers. 
A donor forms four covalent bonds with the nearby Si atoms, and its core is therefore close to D$^{5+}$ plus the contribution of the four covalent electrons. 
For Bi donors \cite{Feiock1969}, $\gamma(\rm{Bi}^{5+}) = -47.24$ and the effect of the four covalent bonds would be to reduce this value down to $\gamma(\rm{Bi}^{0})=-159$. 
As Bi donor and Sb donors have different numbers of core electrons, the antishielding parameters may be very different.
The determination of the antishielding parameters is beyond the scope of this work. 
Another hypothesis is the presence of several defects in the vicinity of a donor, leading to an underestimation of the orbital mixing. 
In summary, we have observed shifts greater than a megahertz of NMR transitions for Bi and Sb donors in Si, and their probable link to NQI. 
The molecular orbital model used in this study points at the mixing of the donor orbital ground states A$_1$ with the E doublet as a possible origin of the EFG. 
As a similar orbital state mixing can also be achieved with an electric field \cite{Pica2014}, or with a global \cite{Franke2016} or a local \cite{Pla2016} strain, the presented results will allow to relate, in the context of devices based on donors in silicon, the device physical properties (local strain, electric potentials defined by local gates) to the magnetic resonance properties of the donor spins. 
Finally, our results imply that by using local gates to modify and control the donor wavefunction, it would be possible to generate on-demand a local EFG at the donor nucleus which could be used to manipulate the donor nuclear spin.

\ack
The authors thank J. J. Pla for discussions. 
The work at Keio has been supported by JSPS KAKEN (S), JSPS Core-to-Core, and 
Spintronics Research Network of Japan. 
GP was funded by the joint EPSRC (EP/I035536) / NSF (DMR-1107606) Materials World Network grant. 
We also acknowledge access to NCRIS facilities (ANFF and the Heavy Ion Accelerator Capability) at the Australian National University.



\providecommand{\newblock}{}

\end{document}